\documentclass[aps,pra,showpacs]{revtex4}
\usepackage{amsfonts,amsmath}
\usepackage[dvips]{graphics}
\usepackage{amsmath,amssymb,amsthm}

\newcommand{\ket}[1]{\mathop{\left|#1\right>}\nolimits}       % |ket>
\newcommand{\Tr}[1]{\mathop{{\mathrm{Tr}}_{#1}}}              % Trace

\newcommand{\ketbra}[2]{| #1\rangle\!\langle #2 |}
\newcommand{\nn}{\nonumber}

\newcommand{\bbR}{{\mathbb R}}
\newcommand{\bbO}{{\mathbb O}}

\newcommand{\ahead}[2]{\genfrac{}{}{0pt}{}{#1}{#2}}

\begin{document}

\title{Eavesdropping of quantum communication from a non-inertial frame}

\author{K. Br\'adler}
\affiliation{Instituto de F\'isica, Universidad Nacional
Aut\'onoma de M\'exico, Apdo. Postal 20-364, M\'exico D.F. 01000}

\email{kbradler@epot.cz}

\date{\today}

\begin{abstract}
We introduce a relativistic version of quantum encryption protocol
by considering two inertial observers who wish to securely
transmit quantum information encoded in a free scalar quantum
field state forming Minkowski particles. In a non-relativistic
setting a certain amount of shared classical resources is
necessary to perfectly encrypt the state. We show that in the case
of a uniformly accelerated eavesdropper the communicating parties
need to share (asymptotically in the limit of infinite
acceleration) just half of the classical resources.
\end{abstract}


\maketitle

Only relatively a small fraction of works on quantum information
theory studies its various aspects in a relativistic
setting~\cite{relativity}. The violation of Bell's inequality for
a quantum entangled state shared by two observers who are not at
rest with respect to each other in special~\cite{spec_bell} or
general relativity~\cite{gen_bell}  attracted a bit more
attention. Also, it was observed that the entanglement is a
quantity depending on a relative acceleration of one of the
observers who, before being accelerated, shared a maximally
entangled bosonic~\cite{alice} or fermionic pair~\cite{dirac}. In
the latter it was found that the entanglement does not vanish even
in the limit of infinite acceleration. The physical effect behind
the scene responsible for this behavior is the Unruh
effect~\cite{unruh} which can be generalized into an arbitrary
spacetime dimension by the thermalization theorem~\cite{wp2}. As a
seminal example of its consequences on the effectivity of quantum
information protocols in a relativistic setting let us recall a
relativistic version of the quantum teleportation
protocol~\cite{teleport}. In this case Alice is a stationary
observer sharing a maximally entangled state with Bob in the same
frame. Later, Bob uniformly accelerates with respect to Alice and
it is found that the fidelity of an input and output state is a
decreasing function of Bob's acceleration. Due to the Unruh
effect, Bob has no longer access to the maximally entangled state
and the higher acceleration he perceives the less entangled state
is effectively shared by both participants. Consequently, the
reliability of the teleportation is reduced. Therefore, this quite
different behavior from the `standard case' leads us to the
revision of the other quantum communication protocols.

In this paper we consider relativistic effects in a cryptographic
protocol known as Private Quantum Channel (PQC). We are interested
in learning how can relativistic effects be used to relax the
assumptions on classical resources needed for the perfect
encryption of a quantum state. It is known that for an
eavesdropping in the same frame where the communication takes
place two bits of shared classical information are necessary for
the secure transmission of the state. We have shown that the
amount of the shared classical resources is reduced if the
eavesdropper (Eve) is in a relative uniform acceleration with
respect to inertial Alice and Bob. In the limit of infinite
acceleration, the classical cost of the secure encryption is
reduced to one half. The infinite acceleration limit corresponds
to a situation in which freely falling Alice and Bob (into a black
hole) transmit information while Eve escaping the fall tries to
tap the communication. Note that in the following analysis we will
encrypt qubits but our results can be directly generalized to the
case of $d$-level states (qudits). On the other hand, the
encryption of states with infinitely many degrees of freedom (so
called continuous-variable quantum states) requires much more
caution even in the non-relativistic case~\cite{CVPQC}.

The structure of the paper is the following. In
Sec.~\ref{sec_packets} we remind the concept of Minkowski
(Rindler) particles as a free massless scalar field given by the
quantization of a superposition of plane waves forming wave
packets in Minkowski (Rindler) spacetime. We will employ an effect
similar to the Unruh effect (derived for the single photon field
excitations), that is, a decomposition of the Minkowski
one-particle state into the many-particle Rindler state occupying
two causally disconnected regions. In Sec.~\ref{sec_fulpartenc} we
recall some of the basic properties of PQCs together with the
justification of their use. The main part comes in
Sec.~\ref{sec_infoofeve} where we derive an upper bound on
information which a non-inertial eavesdropper, in the limit of
infinite acceleration, is able to extract from partially encrypted
communication being in progress between two legal inertial
observers.

\section{Transformation of Minkowski particles into a non-inertial frame}
\label{sec_packets}

The wave packets, whose properties will be employed, were studied
in detail~in~\cite{wp1} (and originally come from~\cite{hawking})
but for more general approach to the particle/wave packet concept
in QFT we refer to Takagi's noticeable earlier paper~\cite{wp2}.
The used wave packets have the usual form of a superposition of
plane waves in both spacetimes modulated by Fourier coefficients
with a simple structure. The quantization was performed for a free
massless scalar bosonic field and the resulting objects in
Minkowski spacetime are known as Minkowski particles (and
similarly for Rindler spacetime). Such an object is spatially and
temporarily localized and thus allows us to talk about
preparation, manipulation and measurement as known from
non-relativistic quantum mechanics. This is not possible with a
single field excitation (Fock state) as understood by QFT (similar
approach with a different kind of quantized wave packets was quite
recently used for predicting of a decoherence effect of an
entangled pair with one of the qubits perceiving changes in the
gravitational field~\cite{wp3}). How could a communication with
Minkowski particle look like? This is in fact the question of the
information encoding, i.e. defining a logical qubit communication
basis. As we will see, the choice of the encoding has an impact on
the calculation using the field transformation from Minkowski to
Rindler spacetime.

Let the Minkowski observers' basis be composed of two
one-Minkowski particles both in two orthogonal spatial modes (here
indexed as 1,2). This is known as the dual rail encoding used, for
example, in linear-optical quantum computing
schemes~\cite{dual_LOQC} or an experimental realization of quantum
teleportation protocol~\cite{dual_tele}. Then, we can write our
logical qubits as
\begin{equation}\label{encoding}
\{\ket{\bbO}\equiv\ket{0}_1^\mathcal{M}\ket{1}_2^\mathcal{M},
\ket{\openone}\equiv\ket{1}_1^\mathcal{M}\ket{0}_2^\mathcal{M}\},
\end{equation}
where $\ket{0},\ket{1}$ is the Minkowski vacuum and the Minkowski
one-particle, respectively. Similarly as for the quantization of
Fock states the field is quantized in both Minkowski and Rindler
spacetime and the corresponding creation and annihilation
operators (before the quantization being the mode expansion
coefficients) are defined. Using the Klein-Gordon inner product
these two sets of operators are related and the creation and
annihilation operators from one space are expressed as linear
combinations of those from the other space. The coefficients in
these combinations are known as the Bogoliubov particle
coefficients yielding~\cite{wp1} the transformed Minkowski vacuum
(the Rindler vacuum)
\begin{equation}\label{vac_Rindler}
    \ket{0}_1^\mathcal{M}=\prod_{k,l}
    {1\over\cosh{r_k}}\sum_{g=0}^\infty\tanh^g{r_k}\ket{g_{-k,l}}_{1L}\ket{g_{k,l}}_{1R},
\end{equation}
where $\cosh{r_k}=\left(1-e^{-2\pi\omega_kc/a}\right)^{-1/2},\tanh
r_k=e^{-\pi\omega_kc/a}$ with $a$ the Rindler observer's proper
acceleration and e.g. $\ket{g_{k,l}}_{1R}$ is the Rindler
$g-$particle state with integer mode numbers $k,l$ characterizing
the particle in Rindler spacetime in spatial mode 1 and the right
wedge. The Minkowski one-particle state as seen by the Rindler
observer has the form~\cite{wp1}
\begin{equation}\label{wp_rindler}
\ket{1_{\tilde k\tilde
l}}_1^\mathcal{M}=\Big[\prod_{\ahead{k\not=k_t}{l\not=l_t}}{1\over\cosh{r_k}}
\sum_{p=0}^\infty\tanh^p{r_k}\ket{p_{-k,l}}_{1L}\ket{p_{k,l}}_{1R}\Big]\times
{1\over\cosh^2{r_{k_t}}}
\sum_{q=0}^\infty\sqrt{1+q}\tanh^q{r_{k_t}}\ket{(1+q)_{-k_t,l_t}}_{1L}\ket{q_{k_t,l_t}}_{1R},
\end{equation}
where the indexes with tildes indicate the particles in Minkowski
space and $k_t,l_t$ are so called equivalent modes in Rindler
spacetime~\cite{wp1} corresponding to the indexes $\tilde k\tilde
l$. As usual, the Rindler observer has access just to one of the
wedges (in our case to the right one). Generally, we can see the
glimpse of similarity to the transformation of ordinary Fock
states.

Let us see how a density matrix in the dual-rail representation
transforms as a whole. Taking the corresponding products of
Eqs.~(\ref{wp_rindler}) and~(\ref{vac_Rindler}) and tracing over
the left wedge the matrix elements acquire the form of
infinite-dimensional matrices. First, for the diagonal elements we
have
\begin{eqnarray}\label{trace_diag}
   \Tr{L}\left[\ketbra{\bbO}{\bbO}\right]
   &=&\prod_{k,l}{1\over\cosh^2{r_k}}\sum_{g=0}^\infty\tanh^{2g}{r_k}
   \ketbra{g_{k,l}}{g_{k,l}}_{1}\nonumber\\
   &\times&\prod_{\ahead{k\not=k_t}{l\not=l_t}}{1\over\cosh^2{r_k}}\sum_{p=0}^\infty\tanh^{2p}{r_k}
   \ketbra{p_{k,l}}{p_{k,l}}_{2}
   \times{1\over\cosh^4{r_{k_t}}}\sum_{q=0}^\infty(1+q)\tanh^{2q}{r_{k_t}}\ketbra{q_{k_t,l_t}}{q_{k_t,l_t}}_{2},
\end{eqnarray}
and similarly for
$\Tr{L}\left[\ketbra{\openone}{\openone}\right]$. We see that both
matrices are diagonal due to the orthogonality relations valid for
$k,l$ and the spatial modes. After tracing over the left wedge of
the transformed off-diagonal matrix elements we get
\begin{eqnarray}\label{trace_offdiag}
   \Tr{L}\left[\ketbra{\bbO}{\openone}\right]
   &=&\prod_{\ahead{k\not=k_t}{l\not=l_t}}{1\over\cosh^2{r_k}}\sum_{p=0}^\infty\tanh^{2p}{r_k}
   \ketbra{p_{k,l}}{p_{k,l}}_{1}
   \times{1\over\cosh^3{r_{k_t}}}\sum_{q=0}^\infty\sqrt{1+q}\tanh^{2q}{r_{k_t}}\ketbra{(1+q)_{k_t,l_t}}{q_{k_t,l_t}}_{1}\nonumber\\
   &\times&\prod_{\ahead{k\not=k_t}{l\not=l_t}}{1\over\cosh^2{r_k}}\sum_{f=0}^\infty\tanh^{2f}{r_k}
   \ketbra{f_{k,l}}{f_{k,l}}_{2}
   \times{1\over\cosh^3{r_{k_t}}}\sum_{g=0}^\infty\sqrt{1+g}\tanh^{2g}{r_{k_t}}\ketbra{g_{k_t,l_t}}{(1+g)_{k_t,l_t}}_{2}\nonumber\\
   &=& \left(\Tr{L}\left[\ketbra{\openone}{\bbO}\right]\right)^T.
\end{eqnarray}
Thus, both off-diagonal elements from Minkowski spacetime become
off-diagonal matrices in Rindler spacetime and a general Minkowski
density matrix (in the dual-rail representation) is symmetrical
and three-diagonal after the transformation.

\section{Perfect and partial quantum state encryption}
\label{sec_fulpartenc}

Having introduced the concept of localized objects both in
Minkowski and Rindler spacetime we can start generalizing PQC into
the relativistic setting. Let Alice and Bob be two inertial
observers in Minkowski spacetime who share a random and private
string of bits -- key and are connected through a quantum channel.
The bits of the key index unitary operations (to be specified) are
defined in a space spanned by the dual rail basis
vectors~(\ref{encoding}). Hence, Alice receives/prepares an
arbitrary and unknown quantum state (generally a mixed state
written in the dual rail basis~(\ref{encoding})) and applies one
of the unitaries dictated by the key. She sends the modified state
down the channel to Bob who makes the inverse unitary operation.
The task for Alice and Bob is to securely transmit the input
(unencrypted) state and to prevent from a potential eavesdropping
by Eve. She has full access to the quantum channel and is able to
construct any generalized measurement (POVM) that could give her
some information about the state. In our case, we suppose that Eve
is a non-inertial observer so let us investigate the consequences
for the protocol security.

The theory of PQCs learns~\cite{PQC} that for any qubit $\Xi$
holds
\begin{equation}\label{fullPQC}
    \Xi\mapsto{1\over4}\sum_{i=1}^4\sigma_i\Xi\sigma_i^\dagger={\openone\over2}=\varrho,
\end{equation}
where the encryption set composed of the Pauli matrices
$\{\sigma_i\}\rightleftharpoons\{\openone,\sigma_X,\sigma_Y,\sigma_Z\}$
is the usual (but non-unique) choice for the encryption procedure.
No matter the form of $\Xi$ someone without the knowledge which
operation for the encryption $\sigma_i$ was used has no chance to
get any information about the state since in the identity density
matrix there is no information on the input state. This is called
a perfect encryption. We also see that necessary (and also
sufficient~\cite{PQC}) length of the shared key for the encryption
of one qubit are two bits. On the other hand, if we don't adhere
to the rule of not using less than a two-bit key and use, for
instance $\sigma_1,\sigma_2$, i.e. just a one-bit key, we get a
partially encrypted qubit
\begin{equation}\label{partPQC}
    \Xi\mapsto{1\over2}\sum_{j=1}^2\sigma_j\Xi\sigma_j^\dagger
    ={1\over2}\begin{pmatrix}
      1 & f(\Xi_{12},\Xi_{21}) \\
      f(\Xi_{12},\Xi_{21}) & 1 \\
    \end{pmatrix}
    =\tilde\varrho.
\end{equation}
Hence, in the off-diagonal terms of $\tilde\varrho$ there is some
hidden information about $\Xi$ and the state is not secured. So,
assuming the partial encryption a possible inertial eavesdropper
would apparently have a chance to get some information on the
input state. But how is it with Eve as the non-inertial
eavesdropper? Before answering this question let us make a small
detour and say something general about the PQC applicability.

Obviously, PQC is meaningful whenever an eavesdropper can get some
information about density matrix elements by any kind of
generalized measurement. Into this situation falls~(a)~the case
when we have several copies of an unknown state (or generally
classically correlated input and thus we can make a quantum state
estimation) or~(b)~the case when we have one copy of an arbitrary
mixed state $\Xi=\sum p_i\Xi_i$ with given a priori probabilities
$p_i$ and we implement one of possible discrimination techniques
such as unambiguous state discrimination (if applicable), minimum
error discrimination or their advanced
combinations~\cite{discrim}. On the contrary, if one has one copy
of an unknown or even known state then the encryption is
unnecessary. In the first case the state is already encrypted (an
unknown qubit is seen as a mixture of all states in the Bloch
Sphere that is proportional to a maximally mixed state) and in the
latter case one just needs to send classical information about the
state preparation and a high-priced quantum channel is not
required for the state transmission (the question of information
load tradeoff between classical and quantum channel is not
interesting for us now).

For later purposes, let us shortly discuss how the leakage of
information of an encrypted state can be measured. The extractable
information is quantified by the accessible information which is a
maximization over all possible POVMs. However, finding the optimal
POVM is a rather difficult task and analytical solutions are known
only for few cases, mostly under some kind of
symmetry~\cite{helstrom}. In this light, the quantity we will use
is the Holevo information~\cite{holevo_bound}
\begin{equation}\label{holevobound}
    \chi(\{p_i,\varrho_i\})=S(\varrho)-\sum_{i=1}^np_iS(\varrho_i)
\end{equation}
bounding the accessible information from above, where
$\varrho=\sum p_i\varrho_i$ and
$S(\varrho)=-\Tr{}\left[\varrho\log_2\varrho\right]$ is the von
Neumann entropy. The Holevo bound is more feasible to calculate
than the accessible information but it is not generally a tight
bound. Fortunately, in the case we are investigating it is so. To
see it, let us assume that an input qubit is fully encrypted
\begin{equation}\label{holevo_exmpl}
    \Xi=\sum_{i=1}^n p_i\Xi_i\stackrel{PQC}{\longmapsto}\sum_{i=1}^n p_i
    \underbrace{{1\over4}\sum_j^4\sigma_j\Xi_i\sigma_j^\dagger}_{\varrho_i\,\propto\,\openone}
    =\varrho\,(\propto\openone).
\end{equation}
Then, for the best possible discrimination of the states before
encryption ($\Xi_i$) Eve has to distinguish among $\varrho_i$
because the non-encrypted states were transformed onto this
'encoding'. Inserting the encoding into~(\ref{holevobound}) we get
$\chi(\{p_i,\varrho_i\})=0$. Of course, after the encryption Eve
may attempt to distinguish any encoding at her will (and the
Holevo bound does not need to be zero) but whatever discrimination
strategy she chooses at the end of the day she always gets zero
information about the occurrence of the states $\Xi_i$ before the
encryption~\cite{PQC}.

\section{Information gain of an accelerated eavesdropper}
\label{sec_infoofeve}

The purpose of this main part is to show that in the limit of
infinite acceleration an eavesdropper (Eve as the Rindler
observer) cannot get any information from just a partially
encrypted qubit. First, we show that in the infinite limit the
Holevo bound for a perfectly encrypted qubit is the same as a
partially encrypted one. Second, despite the Holevo bound not
being a tight upper bound for the accessible information we show
that the accessible information available to the accelerated
eavesdropper tapping a partially encrypted qubit is zero as well.
We may write
\begin{widetext}
\begin{align}\label{entropy}
\left|\chi(\{p_i,\varrho_i\})-\chi(\{\tilde
p_i,\tilde\varrho_i\})\right| & =
\left|S(\varrho)-S(\tilde\varrho)
    +\sum_{i=1}^n\tilde p_iS(\tilde\varrho_i)-\sum_{i=1}^np_iS(\varrho_i)\right|\nn\\
& \leq |S(\tilde\varrho)-S(\varrho)|
    +\left|\sum_{i=1}^n\tilde p_iS(\tilde\varrho_i)-\sum_{i=1}^np_iS(\varrho_i)\right|
    && \text{by  the triangle inequality.}
\end{align}
\end{widetext}
Next, in the spirit of Eq.~(\ref{holevo_exmpl}) we use
$S(\varrho_i)=S(\varrho)$ and prove the convergence of the second
summand in the second row of Eq.~(\ref{entropy}). Later, using the
same equality, we show that the convergence of the first summand
follows from a simple modification of the proof of the convergence
of the second summand. It is a special case simply by putting
$n=1\ (\tilde p_i=1)$, i.e. omitting the sum over $n$ in all the
derivations which follow.

Hence, rewriting the second summand in Eq.~(\ref{entropy}) and
generating its Taylor series about any (identical) nonzero point
we get after the rearranging~\footnote{We are basically looking
for the Taylor series of a multivariate function which is in the
form $g(x_1,\dots,x_{n+1})=\sum_{i=1}^{n+1}g(x_i)$. Inserting this
function into the prescription for the Taylor
series~\cite{taylor_multi} we find that it corresponds to the
summation of the Taylor decompositions of particular $g(x_i)$
(generally about different points). In our case we put everywhere
the same point.}
\begin{align}\label{entropy_eigenval}
    \sum_{j=1}^\infty\left(
    \sum_{i=1}^n\tilde p_i\tilde\lambda_{ij}\log_2\tilde\lambda_{ij}
    -\lambda_j\log_2\lambda_j\right)
    =\sum_{j=1}^\infty\left(\sum_{k=0}^\infty
    \left(b_k\sum_{i=1}^n\tilde
    p_i\tilde\lambda_{ij}^k-\lambda_j^k\right)\right)
    =\sum_{k=0}^\infty\left( b_k
      \sum_{i=1}^n\tilde
      p_i\Tr{}\left[\tilde\varrho_i^k\right]-\Tr{}\left[\varrho^k\right]\right),
\end{align}
where $\tilde\lambda_{ij},\lambda_j$ is the $j-$th eigenvalue of
$\tilde\varrho_i$ and $\varrho$, respectively, and
$\sum_{j=1}^\infty\tilde\lambda_{ij}^k=\Tr{}\left[\tilde\varrho_i^k\right],
\sum_{j=1}^\infty\lambda_{j}^k=\Tr{}\left[\varrho^k\right]$ was
used. $b_k$ are the Taylor expansion coefficients and are set up
to be identical for all expanded functions in our
case~[\arabic{footnote}]. Eq.~(\ref{entropy_eigenval}) is
trivially equal to zero if $k=0$, the same holds for $k=1$
($\tilde\varrho_i,\varrho$ are density matrices and $\sum \tilde
p_i=1$). For $k=2$ we will show its asymptotic convergence to zero
in Rindler space by a direct calculation. This will be a starting
point for proving the convergence for $k>2$. First, we set up the
following notation
\begin{equation}\label{elements_of_rhotilde}
    a_q={\tanh^{2q}r\over\cosh^4r}(1+q)
    \hspace{.5cm}
    A_q={\tanh^{2q}r\over\cosh^3r}\sqrt{1+q}
\end{equation}
($r\in\bbR^+$) where $a_q$ are the elements of
Eq.~(\ref{trace_diag}) forming the diagonal
of~Eqs.~(\ref{fullPQC}) and~(\ref{partPQC}) in Rindler spacetime.
Similarly, the coefficients $A_q$ coming
from~Eq.~(\ref{trace_offdiag}) are the off-diagonal elements
of~Eq.~(\ref{partPQC}) in Rindler spacetime. The off-diagonal
coefficients of the matrices $\tilde\varrho_i$ (coming from
Eq.~(\ref{partPQC}) for $\Xi=\sum p_i\Xi_i$) are the scalar
functions $f_i(\Xi_{12},\Xi_{21})$ independent of the acceleration
and do not transform. Also, note that we work just with one
mode~$k,l$~\footnote{It is sufficient to work with just one mode
because the matrix elements in Rindler space are mode products of
sums of particular modes. Proving the convergence of the sum just
for one particular mode and using $0\leq a_q,A_q\leq1$ we show
that the whole product converges.}. Next, for the upcoming
considerations we need to know that
$a_q>a_{q+1},\lim_{r\to\infty}\sum_{q=0}^\infty a_q^2=0,0\leq
a_q\leq1$ and the same holds for $A_q$. Writing the square of
$\tilde\varrho_i$ we get a five-diagonal infinite-dimensional
matrix where on the diagonals we generally find sequences of the
type $\sum a_qa_p,\sum A_qA_p$ and $\sum a_qA_p$~\footnote{Except
$a_q,A_q$ we can also find there $\sum_q\tanh^{2q}r/\cosh^2r$ for
which convergence is easy to see.}. Due to the properties of
$a_q,A_q$ all of them can be bounded from above by putting $p=q$
thus the sequences (and of course their elements) converge to zero
for $r\to\infty$.

To continue let us consider the following inequalities
\begin{equation}\label{ineq_trace}
    \sum_{i=1}^n\tilde p_i\Tr{}\left[\tilde\varrho_i^k\right]-\Tr{}\left[\varrho^k\right]
    \leq\sum_{i=1}^n\Tr{}\left[\tilde\varrho_i^k\right]-\Tr{}\left[\varrho^k\right]
    \leq\Tr{}\left[\left(\sum_{i=1}^n\tilde\varrho_i+\varrho\right)^k\right]
    \leq\Tr{}\left[(n+1)^k\tilde\varrho_{\text{max}}^k\right],
\end{equation}
where, as stated, all $\tilde\varrho_i$ have the same form as
Eq.~(\ref{partPQC}) but with different scalar functions
$f_i(\Xi_{12},\Xi_{21})$. Thus, the maximum is taken over $n$
scalar functions. The last inequality in Eq.~(\ref{ineq_trace})
does not hold in general but the structure of
$\tilde\varrho_i,\varrho$ allows us to do it (recall that we are
in Rindler spacetime where $\varrho$ is a diagonal density matrix
and $\tilde\varrho_i$ are three-diagonal density matrices with the
main diagonal equal to the diagonal of $\varrho$). Then
\begin{equation}\label{ineq_trace_limits}
    \lim_{r\to\infty}\left(\sum_{i=1}^n\Tr{}\left[\tilde\varrho_i^k\right]-\Tr{}\left[\varrho^k\right]\right)
    \leq\lim_{r\to\infty}\Tr{}\left[\left(m\tilde\varrho_{\text{max}}^2\right)^{k/2}\right]
    =m\Tr{}\left[\left(\lim_{r\to\infty}\tilde\varrho_{\text{max}}^2\right)^{k^\prime}\right],
\end{equation}
where $m=(n+1)^2,k^\prime=k/2$ (note that for the convergence of
the first summand in Eq.~(\ref{entropy}) we have $m=1$ and thus
there is no need for the maximization over $n$). We use some basic
limit properties in the last equality with
$\lim_{r\to\infty}\tilde\varrho_{\text{max}}^2$ meaning the limit
of {\em every} element of the matrix
$\tilde\varrho_{\text{max}}^2$. Since for $r\to\infty$ the limit
converges to zero its $k^\prime-$th matrix power converge as well
thus proving the convergence of~Eq.~(\ref{entropy_eigenval}) to
zero. Consequently, we bounded the Holevo information
difference~(\ref{entropy}) from above.

Since the Holevo bound is not a tight bound on the accessible
information it remains one step more and we follow up with the
discussion from the end of Sec.~\ref{sec_fulpartenc}. From
Eq.~(\ref{entropy}) we know that the difference goes
asymptotically to zero. Now we choose the encoding
$\{p_i,\varrho_i\}$ of the perfect PQC such that
$\chi(\{p_i,\varrho_i\})=0$ both in Minkowski and Rindler
spacetime. In fact it corresponds to Eve's choice of the
discrimination strategy in Rindler spacetime. Although we cannot
dictate Eve what to do, in this case, due to the properties of
perfect PQC, any tapping strategy she is able to do gives her no
information. Then, from the convergence above together with
Eq.~(\ref{entropy}) it follows that the Rindler observer (Eve)
cannot get any information even from the partially encrypted
quantum state.

\section{Conclusions}

Wa addressed the question of the role of the Unruh effect on
security of quantum communication. We considered the setting where
two honest parties are at rest (Minkowski observers) and an
eavesdropper is uniformly accelerated with respect to them
(Rindler observer). The Minkowski observers established PQC where
logical qubits are represented by a free scalar field in the form
of so-called Minkowski particles considering the dual-rail
encoding. The legal parties did not satisfy the security
requirements and they encrypted a qubit with just one bit of a
classical key instead of two bits. Since this is a necessary and
sufficient condition for the perfect encryption, it would
generally lead to a leakage of some information about the qubit.
However, we have shown that in the limit of infinite acceleration
this partial encryption is sufficient to secure the qubit and,
thus, it is prevented form getting any information into the hands
of an eavesdropper.

In other words, we have shown that the Unruh effect makes the
communication noisy in the direction where the eavesdropper in an
non-inertial frame cannot distinguish between a perfectly and
partially encrypted qubit. The important aspect of the above
conclusion is that we did not consider problematic single photon
excitations (Fock states) but temporarily and spatially localized
objects both in Minkowski and Rindler spacetime (Minkowski and
Rindler particles, respectively). The problem of Fock states lies
in their delocalization making difficult to talk about the
preparation or manipulation (considering the production of Fock
states in a realistic cavity which is subsequently accelerated
brings even more problems). This is the first step toward the
description of a more natural and general setup where both the
legal participants and Eve are differently accelerated and the
objective would be to enumerate the leaked information not only in
the asymptotic case.

\begin{acknowledgments}
The author is very grateful for valuable discussions with
R.~J\'auregui, R. Mann and I. Fuentes-Schuller and G. J. Milburn
for providing paper~\cite{wp3} before its posting.
\end{acknowledgments}

\end{document}